\documentclass{article}
\usepackage{amssymb}
\usepackage{amsmath}
\usepackage{graphicx}

\begin{document}

\begin{center}
\Large\textbf{On concentration dependence of arsenic diffusivity}
\\[2ex]
\normalsize
\end{center}

\begin{center}
\textbf{O. I. Velichko}

\bigskip


{\it E-mail address (Oleg Velichko):} velichkomail@gmail.com
\end{center}

\textit{Abstract.} An analysis of the equations used for modeling
thermal arsenic diffusion in silicon has been carried out. It was
shown that for arsenic diffusion governed by the vacancy-impurity
pairs and the pairs formed due to interaction of impurity atoms
with silicon self-interstitials in a neutral charge state, the
doping process can be described by the Fick's second law equation
with a single effective diffusion coefficient which takes into
account two impurity flows arising due to interaction of arsenic
atoms with vacancies and silicon self-interstitials, respectively.

Arsenic concentration profiles calculated with the use of the
effective diffusivity agree well with experimental data if the
maximal impurity concentration is near the intrinsic carrier
concentration. On the other hand, for higher impurity
concentrations a certain deviation in the local regions of arsenic
distribution is observed. The difference from the experiment can
occur due to the incorrect use of effective diffusivity for the
description of two different impurity flows or due to the
formation of nonuniform distributions of neutral vacancies and
neutral self-interstitials in heavily doped silicon layers.

\section{Introduction}
Arsenic is the basic \textbf{\textit{n}}-type impurity in the
technology of manufacturing silicon semiconductor devices and
integrated microcircuits. It is due to the small difference
between the atomic radii of arsenic and silicon that results in a
high value of arsenic solubility limit in silicon. At present,
arsenic doped regions are formed by ion implantation with
subsequent thermal annealing of implanted layers. The thermal
treatment provides the electric activation of impurity atoms,
recrystallization of amorphous layers, and annealing of radiation
defects created by ion implantation. During annealing a transient
enhanced diffusion of impurity atoms occurs that results in the
increase in the doped region dimensions and, respectively, in the
increase of the \textbf{\textit{p-n}} junction depth. On the other
hand, due to the strong dependence of the arsenic diffusivity on
impurity concentration, there occurs the formation of the specific
impurity distribution of a ``box-like'' form. These impurity
distributions are characterized by a practically constant arsenic
concentration within the limits of the doped region and a very
sharp fall of impurity concentration in the bulk of a
semiconductor. Such kind of impurity atom distribution is very
important in semiconductor manufacturing because it imparts high
electrophysical characteristics to semiconductor devices. It
follows from the presented analysis that knowing the concentration
dependence of arsenic diffusivity is very important for solving
the problems of doping process modeling in the semiconductor
technology. Investigation of different concentration dependences
of arsenic diffusivity is the goal of this research.

\section{Model of diffusion of arsenic atoms}
To investigate the concentration dependence of arsenic
diffusivity, it is reasonable to consider the diffusion processes
which are independent of the factors that complicate the diffusion
description, such as arsenic clustering or precipitation,
annealing of the defects created by ion implantation, oxidation of
the surface of a semiconductor, etc. Therefore, let us consider
thermal arsenic diffusion from a constant dopant source on a
surface that ensures the maximal impurity concentration below a
solubility limit of arsenic in silicon. Let us also assume that
the distributions of neutral point defects responsible for
impurity diffusion are approximately uniform within the diffusion
zone \cite{Velichko-1985} and the local charge neutrality is valid
\cite{Vaskin-67,Shrivastava-80,Velichko-1988,Velichko-Klimovich-88}.
Then, the system of equations for impurity diffusion due to the
formation, migration, and dissociation of the ``impurity atom
--- vacancy'' and ``impurity atom --- silicon self-interstitial''
pairs, being in a local thermodynamic equilibrium with the
substitutionally dissolved impurity atoms and point defects
participating in the pair formation \cite{Velichko-84}, takes the
following form:

\begin{equation} \label{DifEq}
\frac{\partial \, C}{\partial \, t} =\nabla \left\{\, \,
\left[D^{E} \left(\chi \right)+D^{F} \left(\chi \right)\right]\,
\, h\, \left(C,C^{B} \right)\; \nabla C\right\} ,
\end{equation}

\begin{equation} \label{Chi}
\chi =\frac{\left(C-C^{B} \right)+\sqrt{\left(C-C^{B} \right)^{2}
+4n_{i}^{2} } }{2n_{i} } \, ,
\end{equation}

\begin{equation} \label{DEF}
D^{E,\, F} \left(\chi \right)\, =D_{i}^{E,\, F} D^{EC,\, FC} (\chi
)=D_{i}^{E,\, F} \frac{1+\beta _{1}^{E,\, F} \chi + \beta
_{2}^{E,\, F} \chi ^{2} }{1+\beta _{1}^{E,\, F} +\beta _{2}^{E,\,
F} } \,  ,
\end{equation}

\begin{equation} \label{hMulti}
h\, \left(C,C^{B} \right)=1+\frac{C}{\sqrt{\left(C-C^{B}
\right)^{2} +4n_{i}^{2} } }  \, .
\end{equation}

\begin{equation} \label{DEFi}
D_{i}^{E,F} =D_{i}^{E\times ,F\times } +D_{i}^{E-,F-}
+D_{i}^{E2-,F2-} \, ,
\end{equation}

\begin{equation} \label{BT1EF}
\beta _{1}^{E,\, F} =\frac{D_{i}^{E-,\, F-} }{D_{i}^{E\times ,\,
F\times } } \, ,  \qquad  \qquad  \qquad  \qquad \beta _{2}^{E,\,
F} =\frac{D_{i}^{E2-,\, F2-} }{D_{i}^{E\times ,\, F\times } } \, .
\end{equation}

Here $C$ and $C^{B} $ are the concentrations of substitutionally
dissolved arsenic atoms and background impurity of the opposite
type of conductivity, respectively; $\chi $ is the concentration
of electrons $n$, normalized to the intrinsic carrier
concentration $n_{i} $; $h\, \left(C,C^{B} \right)$ is the factor
describing the influence of built-in electric field on drift of
charged pairs; $D_{i}^{E\times }$, $D_{i}^{E-}$, and $D_{i}^{E2-}$
are the partial intrinsic diffusivities of dopant atoms due to
interaction with neutral, singly, and doubly charged vacancies,
respectively; $\beta _{1}^{E}$ and $\beta _{2}^{E}$ are the
empirical constants that describe the relative contribution of
singly and doubly charged vacancies to the arsenic diffusion.
Likewise, $D_{i}^{F\times}$, $D_{i}^{F-}$, and $D_{i}^{F2-}$ are
the partial intrinsic diffusivities of arsenic atoms due to
interaction with neutral, singly, and doubly charged
self-interstitials, respectively; $\beta _{1}^{F}$ and $\beta
_{2}^{F}$ are the empirical constants that describe the relative
contribution of singly and doubly charged self-interstitials to
the impurity diffusion.

It follows from Eq. \eqref{DifEq} and expressions \eqref{DEF},
\eqref{hMulti}, \eqref{DEFi}, and \eqref{BT1EF} that to calculate
the arsenic concentration distribution it is necessary to know the
intrinsic diffusivities $D_{i}^{E} $ and $D_{i}^{F} $ that take
into account the interaction of impurity atoms with vacancies and
self-interstitials, respectively, and 4 values of the empirical
constants $\beta _{1}^{E}$, $\beta _{2}^{E}$, $\beta _{1}^{F}$,
$\beta _{2}^{F}$ that describe the relative contribution of the
charged point defects. Unfortunately, it is very difficult to find
these values because even the values of the relative contribution
of the diffusion mechanism via self-interstitial to the total
arsenic transport

\begin{equation} \label{fi}
f^{I} =\frac{D_{i}^{\, F} }{D_{i} } =\frac{D_{i}^{\, F} }{D_{i}^{E} +D_{i}^{F} }
\end{equation}

\noindent are characterized by well-marked differences as follows
from \cite{Pichler-04}. For example, $f^{I} $ = 0.09 in the
temperature range 975--1200 $^{\circ}$C according to the data of
\cite{Fair-80}, whereas $f^{I} $ = 0.35-0.55 for a temperature of
1000 $^{\circ}$C according to \cite{Ural-99}.

In view of the wide scatter of the values of $f^{I} $ and the
complexity of separate determination of the parameters $\beta
_{1}^{E} $, $\beta _{2}^{E} $, $\beta _{1}^{F} $, and $\beta
_{2}^{F} $, a more simple diffusion equation:

\begin{equation} \label{DifEqEff}
\frac{\partial \, C}{\partial \, t} =\nabla \left\{\, \,
\left[D\left(\chi \right)\right]\, \, h\, \left(C,C^{B} \right)\;
\nabla C\right\} \, ,
\end{equation}

\noindent with the effective diffusivity

\begin{equation} \label{Di}
D\left(\chi \right)\, =D_{i} D^{C} (\chi )=D_{i}
\frac{1+\beta _{1} \chi +\beta _{2} \chi ^{2} }{1+\beta _{1} +\beta _{2} }
\end{equation}

\noindent is very often used for modeling arsenic diffusion. The
empirical constants $\beta _{1}$ and $\beta _{2}$ in expression
\eqref{Di} are found from the best fit to experimental data. For
example, in Ref. \cite{Martinez-Limia-08} a fitting routine based
on different experimental data was used and temperature
dependences of the arsenic partial diffusivities $D_{i}^{\times }
$, $D_{i}^{-} $, $D_{i}^{2-} $ due to neutral, singly, and doubly
charged intrinsic point defects were evaluated, which allows one
to calculate the values of $D_{i} =D_{i}^{\times } +D_{i}^{-}
+D_{i}^{2-} $, $\beta _{1} ={D_{i}^{-}
\mathord{\left/{\vphantom{D_{i}^{-} D_{i}^{\times }
}}\right.\kern-\nulldelimiterspace} D_{i}^{\times } } $ and $\beta
_{2} ={D_{i}^{2-} \mathord{\left/{\vphantom{D_{i}^{2-}
D_{i}^{\times } }}\right.\kern-\nulldelimiterspace} D_{i}^{\times
} } $ for different temperatures of thermal treatment.

It is worth noting that impurity diffusion due to the ``impurity
atom --- silicon self-interstitial'' pairs can occur basically via
interaction with self-interstitials in the neutral charge state.
Then, the description of impurity diffusion by Eq.
\eqref{DifEqEff} and concentration dependence of diffusivity
\eqref{Di} is identical to the use of the initial equation
\eqref{DifEq} and concentration dependence \eqref{DEF}. Really, it
is possible in this case to present the concentration dependence
\eqref{Di} as a sum of concentration dependencies \eqref{DEF}
written for diffusion due to the vacancy-impurity pairs and for
diffusion via the ``impurity atom --- silicon self-interstitial''
pairs:

\begin{equation} \label{DiEq}
D_{i} \frac{1+\beta _{1} \chi +\beta _{2} \chi ^{2} }{1+\beta _{1}
+\beta _{2} } =D_{i}^{E} \frac{1+\beta _{1}^{E} \chi +\beta
_{2}^{E} \chi ^{2} }{1+\beta _{1}^{E} +\beta _{2}^{E} }
+D_{i}^{F\times } \, .
\end{equation}

It follows from Eq. \eqref{DiEq} that

\begin{equation} \label{DiEff}
D_{i} =D_{i}^{E\times } +D_{i}^{F\times } +D_{i}^{E-} +D_{i}^{E2-}
\, ,
\end{equation}

\begin{equation} \label{BT1Eff}
\beta _{1} ={D_{i}^{E-}  \mathord{\left/{\vphantom{D_{i}^{E-}
\left(D_{i}^{E\times } +D_{i}^{F\times } \right)}}\right.\kern-
\nulldelimiterspace} \left(D_{i}^{E\times } +D_{i}^{F\times }
\right)} \, , \qquad  \qquad  \qquad  \qquad \beta _{2}
={D_{i}^{E2-} \mathord{\left/{\vphantom{D_{i}^{E2-}
\left(D_{i}^{E\times } +D_{i}^{F\times }
\right)}}\right.\kern-\nulldelimiterspace} \left(D_{i}^{E\times }
+D_{i}^{F\times } \right)} \, .
\end{equation}

\section{Simulation of arsenic diffusion in silicon}

\begin{figure}[!ht]
\centering {
\begin{minipage}[!ht]{9.4 cm}
{\includegraphics[scale=0.8]{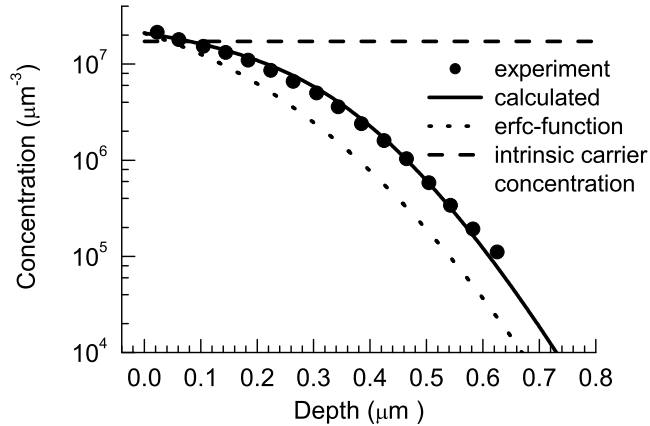}}
\end{minipage}}
\caption{Arsenic concentration profiles formed by thermal
diffusion from a constant source on the silicon surface. The
dotted and solid curves represent the arsenic concentration
profiles calculated from analytical solution \eqref{Cerfc} and
obtained by numerical solution of the diffusion equation that
takes account of the concentration dependence of arsenic
diffusivity and drift of the charged species in the built-in
electric field. Diffusion temperature is 1108 $^{\circ}$C for 120
min. Filled circles are the experimental data of \cite{Chiu-71}}
\label{fig:Chiu-Low}
\end{figure}

In Fig.~\ref{fig:Chiu-Low} the arsenic concentration profile
calculated for the case of thermal diffusion from a constant
source on the surface of a semiconductor is presented. The
experimental data obtained in \cite{Chiu-71} are used for
comparison. In Ref. \cite{Chiu-71}, arsenic diffusion was carried
out at a temperature of 1108 $^{\circ}$C for 120 min. The arsenic
concentration profile was measured by neutron activation analysis.

The analytical solution for the constant value of arsenic
diffusivity, i.e., without taking account of the concentration
dependence $D=D\left(\chi \right)$ and the drift of charged pairs
in built-in electric field, is

\begin{equation} \label{Cerfc}
C(x,t)=C_{S} \, \mathrm{erfc} \left(\frac{x}{L} \right)
\end{equation}

\noindent which is also presented by the dotted curve in
Fig.~\ref{fig:Chiu-Low}. Here $C_{S} ={\rm const}$ is the arsenic
concentration on the surface of a semiconductor, $L=2\,
\sqrt{D_{i} t}$ is the characteristic length of impurity diffusion
and $\mathrm{erfc} \left(x\right)=1-\mathrm{erf} \left(x\right)$
is the additional error function.

The intrinsic diffusivity $D_{i} $ = 2.55$\times$10${}^{-6}$ $\mu
$m${}^{2}$/s and the parameters $\beta _{1} $ = 611 and $\beta
_{2} $ = 44.94 that prescribe a concentration dependence of
impurity diffusivity \eqref{Di} at a temperature of 1108
$^{\circ}$C have been obtained from the expression of
\cite{Martinez-Limia-08}:

\begin{equation} \label{DEffMartinez}
\begin{array}{l} {D^{{\rm As}} \left(\chi \right)\, ={\rm \; \; 4.95}
\times {\rm 10}^{{\rm -4}} \, {\rm exp}\left({\rm -} \displaystyle
\frac {\; 1.4\; {\mathrm{eV}}}{k_{B} T} \right)+{\rm \;
6.103}\times {\rm 10}^{{\rm 8}} \, {\rm exp}\left({\rm -}
\displaystyle \frac {\; {\rm \; 3.95}\; {\mathrm{eV}}}{k_{B} T}
\right)\, \chi {\rm \; }}
\\ {\quad \quad \quad \quad \quad +{\rm 7.76}\times {\rm 10}^{{\rm
11}} \, {\rm exp}\left({\rm -} \displaystyle \frac{\; 5.1\;
{\mathrm{eV}}}{k_{B} T} \right)\, \chi ^{2} \quad \quad \quad
\left[\, {\rm \mu m}^{{\rm 2}} {\rm /s\; }\right]\; }
\end{array} \, .
\end{equation}

It is clearly seen from Fig.~\ref{fig:Chiu-Low} that the account
of the drift of charged pairs $\left({\rm As}^{+} {\rm D}^{\times
} \right)$ and $\left({\rm As}^{+} {\rm D}^{2-} \right)$ and also
the account of the concentration dependence of arsenic diffusivity
due to the increased concentrations of charged defects ${\rm
D}^{-} $ and ${\rm D}^{2-} $ in the region with higher impurity
concentration results in a good agreement of the calculated
profile with experimental data. It is worth noting that the
analytical solution gives smaller concentration values and,
respectively, smaller depth of the \textbf{\textit{p-n}} junction
in comparison with the experiment.

\begin{figure}[!ht]
\centering {
\begin{minipage}[!ht]{9.4 cm}
{\includegraphics[scale=0.8]{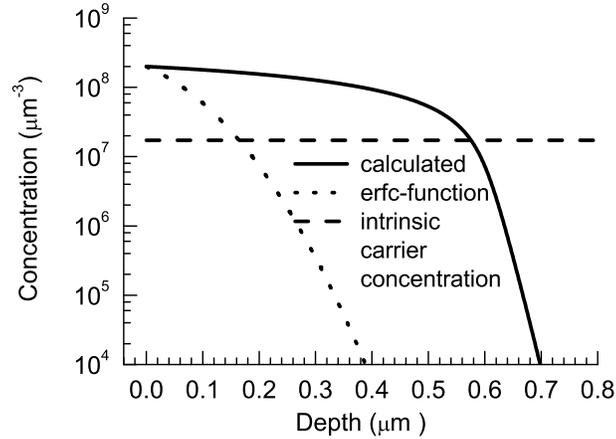}}
\end{minipage}}
\caption{Arsenic concentration profiles formed by thermal
diffusion from a constant source on the silicon surface. The
dotted and solid curves represent the arsenic concentration
profiles calculated from analytical solution \eqref{Cerfc} and
obtained by numerical solution of the diffusion equation that
takes account of the concentration dependence of arsenic
diffusivity and drift of the charged species in the built-in
electric field. Diffusion temperature is 1108 $^{\circ}$C for 30
min} \label{fig:Profiles-mod}
\end{figure}

Based on the good agreement of numerical calculations with
experimental data \cite{Chiu-71}, one can carry out modeling of
arsenic diffusion for higher impurity concentration at the surface
of a semiconductor $C_{S}$ = 2.0$\times$10${}^{8}$ $\mu
$m${}^{-3}$. It follows from Ref. \cite{Solmi-01} that the maximal
value of equilibrium electron concentration $n_{e}$ that can be
obtained in arsenic doped silicon at a temperature of 1108
$^{\circ}$C is equal to 4.24$\times$10$^{8}$ $\mu $m$^{-3}$. A
further increase of electron concentration is limited by the
clustering of impurity atoms. Taking into account these data on
$n_{e}$, it might be expected that for $C_{S}$ =
2.0$\times$10$^{8}$ $\mu $m$^{-3}$ the arsenic clustering is
negligible and can be excluded in diffusion modeling. The arsenic
concentration profile after thermal treatment for 30 min
calculated under these assumptions is presented in
Fig.~\ref{fig:Profiles-mod}.

It can be seen from Fig.~\ref{fig:Profiles-mod} that the form of
arsenic profile differs substantially from the distribution
presented in Fig.~\ref{fig:Chiu-Low} and from analytical solution
\eqref{Cerfc}. Indeed, due to the significant increase in arsenic
diffusion in the high concentration region, the impurity
distribution in this part of the profile becomes flat and differs
substantially from the ${\rm erfc}$-function. On the other hand,
the arsenic profile becomes abruptly falling in the bulk of a
semiconductor and it is characterized by a significant
concentration gradient. It means that using a strong nonlinear
dependence of arsenic diffusivity on impurity concentration, it is
possible to form concentration profiles of electrically active
arsenic atoms with an ideal ``box-like'' form.

\begin{figure}[!ht]
\centering {
\begin{minipage}[!ht]{9.4 cm}
{\includegraphics[scale=0.8]{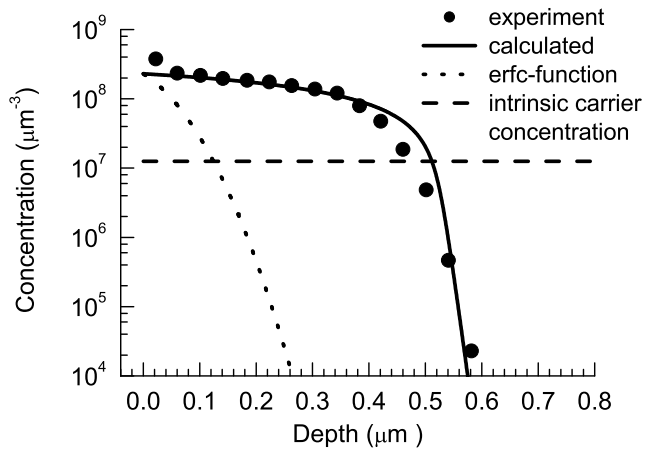}}
\end{minipage}}
\caption{Arsenic concentration profiles formed by thermal
diffusion from a constant source on the silicon surface. The
dotted and solid curves represent the arsenic concentration
profiles calculated from analytical solution \eqref{Cerfc} and
obtained by numerical solution of the diffusion equation that
takes account of the concentration dependence of arsenic
diffusivity and drift of the charged species in the built-in
electric field. Diffusion temperature is 1050 $^{\circ}$C for 60
min. Filled circles are the experimental data of \cite{Chiu-71}}
\label{fig:Profiles-High}
\end{figure}

Let us compare now the calculations of diffusion process
characterized by a strong nonlinear concentration dependence of
diffusivity with experimental data. Figure~\ref{fig:Profiles-High}
presents the arsenic concentration profile calculated for the
treatment temperature of 1050 $^{\circ}$C and duration of 60
minutes. The parameters of arsenic diffusivity $D_{i} $ =
5.8134$\times$10${}^{-7}$ $\mu $m${}^{2}$/s; $\beta _{1}$ = 238.9;
and $\beta _{2} $ = 12.7 were taken from \cite{Martinez-Limia-08}.
The value of the impurity concentration at the surface of a
semiconductor $C_{S} $ has been chosen equal to
2.3$\times$10${}^{8}$ $\mu $m${}^{-3}$. This value is below the
maximal equilibrium electron concentration value $n_{e}$ =
3.566$\times$10${}^{8}$ $\mu $m${}^{-}$${}^{3}$ for the annealing
temperature under consideration \cite{Solmi-01}. Therefore, it is
possible to assume that arsenic clustering does not play an
important role in diffusion.

\begin{figure}[!ht]
\centering {
\begin{minipage}[!ht]{9.4 cm}
{\includegraphics[scale=0.8]{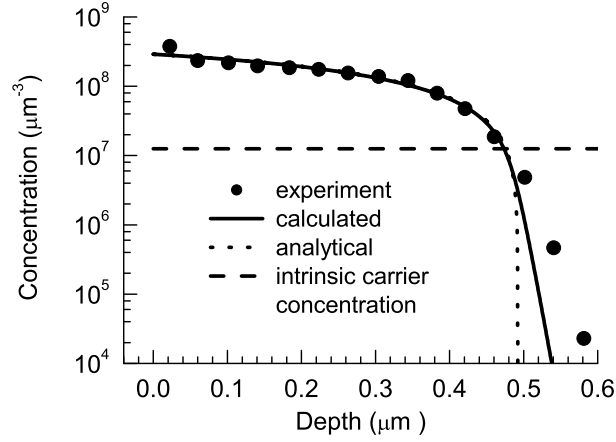}}
\end{minipage}}
\caption{Arsenic concentration profiles formed by thermal
diffusion from a constant source on the silicon surface. The
dotted and solid curves represent the arsenic concentration
profiles calculated from analytical solution \eqref{CTsai} and
obtained by numerical solution of the diffusion equation.
Diffusion temperature is 1050 $^{\circ}$C for 60 min. Filled
circles are the experimental data of \cite{Chiu-71}}
\label{fig:Profiles-Tsai}
\end{figure}

It can be seen from Fig.~\ref{fig:Profiles-High} that the
calculated arsenic concentration profile agrees well with
experimental data both in the region of the high impurity
concentration and in the region of the low concentration of
impurity atoms including the region of \textbf{\textit{p-n}}
junction. On the other hand, some difference between the
calculated and experimental profiles is observed in the region
where a strong decrease in the impurity concentration begins. For
comparison, in Fig.~\ref{fig:Profiles-Tsai} the results of
simulation of arsenic diffusion based on the model of
\cite{Tsai-80} are presented. It is supposed in this model that
doubly charged point defects do not significantly contribute to
arsenic diffusion. The following parameters that describe arsenic
diffusion have been used in simulation: $D_{i} $ =
5.537$\times$10${}^{-7}$ $\mu $m${}^{2}$/s, $\beta _{1} $ = 100,
and $\beta _{2} $ =0 \cite{Tsai-80}. The impurity concentration at
the surface $C_{S} $ has been chosen equal to
2.9$\times$10${}^{8}$ $\mu $m${}^{-}$${}^{3}$.

It can be seen from Fig.~\ref{fig:Profiles-Tsai} that both the
arsenic concentration profiles obtained by analytical solution and
those by numerical calculation agree well with experimental data
excluding the region of a low impurity concentration. It is worth
noting that here the approximate analytical solution of the
diffusion equation \eqref{DifEqEff} for $\beta _{2} $ =0 obtained
in \cite{Nakajima-71} was used:

\begin{equation} \label{CTsai}
C(x,t)=C_{S} \left[1.00-0.87\; \left(\frac{u}{u_{0} }
\right)-0.45\, \left(\frac{u}{u_{0} } \right)^{2} \right] \, ,
\end{equation}

\noindent where

\begin{equation} \label{UTsai}
u=\displaystyle \frac{x}{\sqrt{t} } \, ,  \qquad  \qquad  \qquad
\qquad u_{0} =2\sqrt{\displaystyle \frac{2D_{i} C_{S} }{n_{i} } }
\, .
\end{equation}

It was assumed in \cite{Nakajima-71} that the concentration
dependence of effective diffusivity is described by the following
expression:

\begin{equation} \label{DTsai}
D^{C} \left(\chi \right)\, \, h\, \left(C,C^{B} \right)=2\;
\left(\frac{C}{n_{i} } \right)^{q} \, .
\end{equation}

\section{Analysis of the results obtained by simulation of impurity diffusion with
different concentration dependencies of arsenic diffusivity}

The arsenic concentration profile presented in
Fig.~\ref{fig:Chiu-Low} shows that the concentration dependence of
arsenic diffusivity obtained in \cite{Martinez-Limia-08} provides
excellent agreement with experimental data on impurity
concentration closed to the value of $n_{i} $. On the other hand,
if the arsenic concentration $C>>n_{i}$ and the concentration
profile of electrically active arsenic gets a ``box-like'' form
(see Fig.~\ref{fig:Profiles-High}), a difference from the
experimental data is observed in the local region where a strong
decrease in the impurity concentration begins. Therefore, for
solving the problems arising in modeling silicon doping with
arsenic it seems more convenient to use the concentration
dependence proposed in \cite{Tsai-80} which is characterized by
neglecting the interaction of impurity atoms with doubly charged
point defects. Indeed, in this case, a good agreement with
experiment is observed in the whole region of high impurity
concentration and only a small difference takes place in the
region of low arsenic concentration which is less important for
subsequent calculations of electrophysical parameters. Taking into
account the results of \cite{Velichko-12}, one can suppose that
this difference appears due to the overlooked possibility of
direct migration of arsenic interstitial atoms.

Incomplete agreement of the calculated arsenic profile with the
experimental ones can also be due to a certain inadequacy of using
Eq. \eqref{DifEqEff} in a heavily doped region $C>>n_{i} $,
instead of a more exact equation \eqref{DifEq}. Indeed, with
increase in the impurity concentration, the concentration of
electrons becomes greater than $n_{i} $. As a result, the
concentrations of negatively charged silicon interstitials are
also increasing. If these defects play an important role in the
impurity transport, the description of diffusion based on Eq.
\eqref{DifEqEff} becomes incomplete and it is necessary to use
more exact equation \eqref{DifEq}. However, there can exist
another explanation of the difference between numerical
calculations and experimental data. Indeed, increase in the
concentration of substitutionally dissolved impurity atoms as well
as increase in the electron concentration can result in the
formation of nonuniform distributions of neutral vacancies and
silicon self-interstitials \cite{Velichko-Fedotov-03}. In this
case, both Eq. \eqref{DifEqEff} and Eq. \eqref{DifEq} give
inadequate description of the impurity diffusion. The solution of
this problem requires further investigations.

\section{Conclusions}
The analysis of the equations used for modeling thermal arsenic
diffusion in silicon has been carried out. With this purpose,
different concentration dependencies of impurity diffusivity that
are used for modeling arsenic transports processes have been
analyzed. It is supposed that diffusion is carried out due to the
formation, migration, and dissociation of ``impurity atom ---
vacancy'' and ``impurity atom --- silicon self-interstitial''
pairs which are being in a local thermodynamic equilibrium with
substitutionally dissolved impurity atoms and point defects
participating in pair formation. It is shown that for arsenic
diffusion governed by the vacancy--impurity pairs and by the pairs
formed due to the interaction of impurity atoms with silicon
self-interstitials in a neutral charge state, the doping process
can be described by the Fick's second law equation with a single
effective diffusion coefficient which takes into account two
impurity flows arising as a result of the interaction of arsenic
atoms with vacancies and silicon interstitials, respectively.

Arsenic concentration profiles calculated with the use of the
effective diffusivity taking into account two different impurity
flows agree well with experimental data if the maximal impurity
concentration is close to the intrinsic carrier concentration
$n_{i}$. On the other hand, similar calculations for impurity
concentrations above $n_{i}$ are characterized by a certain
deviation from experimental data in local regions of arsenic
distribution. So, the use of a diffusion model that takes account
of the significant contribution of doubly negatively charged point
defects leads to the deviation in the region where a significant
decrease in the impurity concentration begins. On the other hand,
the neglect of the contribution of these defects leads to a
difference in the region of a low impurity concentration. This
difference can be due to the incorrectness which arises if a
single effective diffusivity is used for the description of two
different impurity flows. On the other hand, the origin of these
small differences can be the formation of nonuniform distributions
of neutral vacancies and neutral silicon self-interstitials in
heavily doped layers.

\end{document}